\documentclass[aps,preprint,nofootinbib]{revtex4}
\usepackage{amsfonts}
\usepackage{amsmath}
\usepackage{amssymb}
\usepackage{graphicx}%
\providecommand{\U}[1]{\protect\rule{.1in}{.1in}}

\begin{document}
\preprint{ }
\title[ ]{Annihilation explosions in macroscopic polyelectrons. Photon
detonation.}
\author{Alexei M. Frolov}
\email{afrolov@uwo.ca}
\affiliation{Department of Chemistry, University of Western Ontario, London,
Canada}

\begin{abstract}
Annihilation of the electron-positron pairs in macroscopic polyelectrons is
considered. It is shown that very fast collapse of the spatial area occupied
by macroscopic polyelectron (or dense electron-positron plasma) produces an
instant annihilation of a very large number of electron-positron pairs. This
phenomenon corresponds to the so-called annihilation explosion. Annihilation
of each electron-positron pair is a highly exothermic process. Therefore, in
dense electron-positron plasma one can observe a very interesting phenomenon
of photon detonation, i.e. a self-organized formation and propagation of the
detonation wave which coincides with the annihilation wave. The photon
detonation can be used in many applications, including many military and
astrophysical problems.
\end{abstract}

\volumeyear{ }
\volumenumber{ }
\issuenumber{ }
\eid{ }
\maketitle

\section{Introduction}

In our previous work \cite{Fro1} we have considered annihilation of the
electron-positron pairs in the three-body Ps$^-$ (= $e^- e^+ e^- = e^{-}_2
e^+$) ion and four-body bi-positronium `molecule' Ps$_2$ ($e^- e^+ e^- e^+ =
e^{-}_2 e^{+}_2$). The main result obtained in \cite{Fro1} indicates clearly
that annihilation of the electron-positron pairs in these two systems
proceeds mainly with the emission of the two and three photons. Briefly,
this means that the annihilation equations in the Ps$^-$ ion and Ps$_2$
`molecule' can be written in one of the two following forms
\begin{equation}
 e^{-} + e^{+} = \hbar \omega_1 + \hbar \omega_2 \; \; \; , \; \; \; or \;
 \; \; e^{-} + e^{+} = \hbar \omega_1 + \hbar \omega_2 + \hbar \omega_3
\end{equation}
Other annihilation channels, e.g., four-photon and five-photon
annihilations, can be ignored in the first approximation which appears to be
remarkably accurate. Moreover, these results for the Ps$^-$ ion and Ps$_2$
system can be generalized to arbitrary polyelectrons, including macroscopic
polyelectrons. In other words, for an arbitrary polyelectron we can also
consider only the two- and three-photon annihilation of the
electron-positron pairs. This allows one to study the electron-positron
plasmas with possible annihilation of particles, since the system of
governing equations is now written in the closed form. The approximate
analytical and accurate numerical solutions of this system of equations can
be found by applying a number of well-known procedures (see, e.g.,
\cite{Kor}, \cite{Num}).

In general, the solution of the governing equations for electron-positron
plasma depends upon the initial and boundary conditions which exist in
reality. It is clear $a$ $priori$ that electron-positron annihilation is
accelerated when the density of electron-positron plasma increases. At very
large densities of such plasma annihilation of the electron-positron pairs
proceeds instantly. This phenomenon is called the photon explosion (or
annihilation explosion). At relatively large densities of the
electron-positron plasma one can observe a very interesting phenomenon of
photon detonation. In this study by photon detonation we mean a
self-organized formation and propagation of the detonation wave in the
dense macroscopic polyelectrons. In this detonation wave the density of the
electron-positron plasma increases suddenly to very large values. This
substantially accelerates annihilation of the electron-positron pairs.
In other words, the detonation wave in dense electron-positron plasmas
always propagates as an annihilation wave.

Our main goal below is to consider annihilation of the $(e^{-},e^{+})-$pairs
in macroscopic polyelectrons. In this work the macroscopic polyelectron (or
polylepton) designates the electrically neutral (or quasi-neutral)
macroscopic system which contains approximately equal (and very large)
numbers of electrons and positrons. In general, an arbitrary macroscopic
polyelectron can be designated as $e^{-}_n e^{+}_m$, where $n \approx m
\approx N_A$ and $N_A$ is the Avogadro number. Below, in this study all
macroscopic polyelectrons are assumed to be neutral, i.e. $m = n$, or
quasi-neutral, i.e. $m \approx n$ and $| m - n | \ll \min (m, n)$.
Macroscopic polyelectrons have a large number of unique properties since
such systems represent a different, non-Born-Oppenheimer world. In this
study we restrict ourselves only to the analysis of annihilation processes
in macroscopic polyelectrons. It should be emphasized that boundness of
macroscopic polyelectrons is not crucial for the goals of this study.
Below, we shall assume that a large number of electrons and positrons is
somehow confined in one spatial area. There is a number of approaches, e.g.,
radiative or magnetic ablation, which can be used to achieve such a
confinement in reality. Note also that our definition of polyelectrons also
includes the case of the confined electron-positron plasma.

\section{Photon detonation in macroscopic polyelectrons.}

In general, annihilation of the electron-positron pairs is a highly
exothermic process. In fact, the energy $E_a$ released during annihilation
of one $(e^{-}, e^{+})-$pair is $E_a \ge 2 m_e c^2 \approx 1.022$ $MeV$. It
follows from here that the energy released per unit mass of the annihilating
electron-positron mixture is quite comparable with the analogous energy
released during nuclear fission and/or thermonuclear burning of the 1:1
deuterium-tritium mixture (below, DT-mixture). An accurate evaluation shows
that the total annihilation of one gram of the electron-positron (1:1)
mixture produces the energy $\approx 4.93 \cdot 10^{10}$ $J$. The same
amount of energy can be obtained from thermal explosion of 11.8 tonnes of
TNT. An analogous amount of thermal energy released during the complete
fission of all nuclei from one gram of Pu-239 is $\approx$ 16.4 tonnes TNT.
Thermonuclear burning of one gram of the 1:1 deuterium-tritium mixture
releases at least 13.8 tonnes TNT.

Bearing this in mind and by applying the well known Khariton's theorem (see,
e.g., \cite{Feok} and references therein) one finds that all macroscopic
polyelectrons are able to detonate in respect to the annihilation, if their
spatial radii exceed some minimal (or critical) value. In a slightly more
general form, we can say that detonation can be achieved, if the product $x$
of spatial radius and electron-positron density, i.e. $x = \rho r$, is
relatively large. In special literature, the value $x = \rho r$ is called
the burn-up parameter. Therefore, the detonation criterion can also be
formulated in the form $x \ge x_{cr}$, where $x_{cr}$ is the critical value
of the burn-up parameter for the electron-positron plasma with density
$\rho_0$. In reality, the minimal spatial radius can be extremely large,
e.g., a few dozens of kilometers. Note that there is a principal difference
between the annihilation explosion (or photon explosion) and photon
detonation. Annihilation explosion can always be observed, e.g., if the
density of annihilating matter is increased to the infinity in a `very
short' time $t_e$. In other words, the photon explosion will proceed in such
cases, when the density of electron-positron macroscopic mixture (i.e.
polyelectron) increases to the infinity with time $t$ as
\begin{equation}
 \rho(t) = \frac{A}{(t - t_e)^n}
\end{equation}
where $A$ and $n$ are some positive constants (in all actual cases $n > 1$).
The explosion time $t_e$ is assumed to be short in comparison with the
life-time of the electron-positron mixture. It should be mentioned that in
contrast with the photon (or annihilation) explosion, the photon detonation
is a self-organized motion which may arise in any annihilating matter, e.g.,
in the macroscopic electron-positron mixture. In fact, the regions of high
density of the electron-positron plasma can also be created during such a
motion. Annihilation of electron-positron pairs in such high-dense areas
proceeds instantly.

Below, we restrict ourselves to the consideration of the photon detonation
only, since this phenomenon is of great interest in numerous applications
as well as for the future theoretical development. A general theory of
photon explosions will be discussed elsewhere. In general, the photon
detonation wave propagating in dense electron-positron mixtures must be a
very intense source of $X-$ray radiation. Note also that the annihilation of
the $(e^-, e^+)-$pairs in macroscopic volumes does not require any minimal
critical density and/or threshold temperature for its ignition. On the other
hand, the rate of energy release in any system undergoing annihilation
rapidly increases at high compressions, since the reaction probability is
proportional to the expectation value of electron-positron delta-function
$\langle \delta_{+-} \rangle$. The proportionality of the overall reaction
rate to the expectation value of the two-particle delta-function is also
true for all working systems based on nuclear fusion. It is also clear that
at high densities the photons emitted during the annihilation of the $(e^-,
e^+)-$pair have significantly better probabilities to redeposit their energy
into surrounding electrons, positrons and atoms. The condition for
`sufficient redeposition' of energy allows one to derive the rigorous
criterion of ignition (or ignition criterion).

Formally, the ignition criterion (or photon detonation criterion) can be
written in the form $L \ge \lambda_R$, where $\lambda_R$ is the Rosseland
mean free path of the annihilation photon and $L$ is the minimal spatial
dimension of the electron-positron mixture (or plasma). The physical
meaning of this condition ($L \ge \lambda_R$) is very transparent:
annihilation photons must leave a substantial part of their energy inside of
the igniting electron-positron mixture. In fact, such an ignition criterion
is only a necessary condition. For actual ignition it must be $L \gg
\lambda_R$. By introducing the Rosseland mean opacity $R_0 =
\frac{1}{\lambda_R \rho}$, where $\rho$ is the macroscopic density of the
electron-positron mixture, one can re-write the ignition criterion in the
form $L \gg \frac{1}{\rho R_o}$. In general, the Rosseland mean opacity
rapidly increases with the density of the plasma. At small densities $\rho
\le 1 \cdot 10^{-3}$ $g \cdot cm^{-3}$ the numerical values of $R_0$ are
very small. Therefore, to decrease the critial size of the annihilating
electron-positron mixture one needs to compress this mixture to very large
densities $\rho$.

Let us write the equations of motion and energy equations for the
electron-positron plasma which also includes radiation transfer. At this
stage of our analysis we shall ignore any presence of atoms, ions and bare
nuclei. Below, the notations $\rho_{-}$ and $\rho_{+}$ stand for the
densities of electrons and positrons, respectively. The total density of
polylepton $\rho_{-} + \rho_{+}$ is designated by $\rho$. The value defined
by the equation $v = \frac{1}{\rho}$ is the specific volume. To simplify the
following analysis we restrict ourselves to the case of planar geometry. In
planar geometry the elementary mass of the macroscopic polyelectron (or
electron-positron plasma) is $dm = \rho dx$. The velocity $v$ of the moving
part of polylepton is $u = \frac{\partial x}{\partial t}$. In this notation
the equation of motion takes the form \cite{Fra}
\begin{equation}
 \frac{\partial u}{\partial t} = - \frac{1}{\rho} \frac{\partial (P +
 q)}{\partial x} = - \frac{\partial (P + q)}{\partial m} \label{e2X}
\end{equation}
where $P = P_e + P_r = P_{-} + P_{+} + P_r$ is the total pressure and $q$
is an additional pressure related with the artificial viscosity (Von-Neuman
artificial viscosity). The notation $P_r$ is the radiation pressure. The
electron ($P_{-}$) and positron ($P_{+}$) pressures are easy related to the
corresponding temperatures, since
\begin{equation}
 P_{-} = k \Bigl(\frac{\rho_{-}}{m_e}\Bigr) T_{-} \; \; \; , \; \; \;
 P_{+} = k \Bigl(\frac{\rho_{+}}{m_e}\Bigr) T_{+}
\end{equation}
where $k$ is the Boltzmann constant and $m_e$ is the electron mass. Below,
we shall assume that $T_{-} = T_{+} = T_e$. Also by assuming that all
photons achieve the Planckian equilibrium with the temperature $T_r$ one
finds the following expression for the radiation pressure $P_r = \frac{4
\sigma}{3 c} T^4_r$, where $\sigma$ is Stephan-Boltzmann constant and $c$ is
the speed of light. In general, the electron and radiation temperatures
differ from each other, i.e. $T_e \ne T_r$.

Let us describe the following two-temperature code which is based on the
radiation diffusion model. In the case of planar geometry the coupled energy
equations take the form
\begin{eqnarray}
 \frac{\partial T_e}{\partial t} = \frac{1}{C_{ve}} \Bigl[ v
 \frac{\partial}{\partial x} \Bigl( K_e \frac{\partial T_e}{\partial x}
 \Bigr) - A_{er} (T_e - T_r) + \frac{dS_e}{dt} - \Bigl( P_e + C_{ve}
 \frac{\partial T_e}{\partial v} \Bigr) \frac{\partial v}{\partial t}
 \Bigr] \label{e4X} \\
 \frac{\partial T_r}{\partial t} = \frac{1}{C_{vr}} \Bigl[ v
 \frac{\partial}{\partial x} \Bigl( K_r \frac{\partial T_r}{\partial x}
 \Bigr) + A_{er} (T_e - T_r) - \Bigl( P_r + C_{vr} \frac{\partial
 T_r}{\partial v} + \frac{4 \sigma}{c} T^4_r \Bigr) \frac{\partial
 v}{\partial t} \Bigr]  ,  \label{e5X}
\end{eqnarray}
where $C_{ve} = \frac32 \frac{k}{m_e}$ and $C_{vr} = \frac{16 \sigma}{c
\rho} T^3_r$ are the specific heats for electron-positron plasma and
radiation, respectively. Also in these equations the corresponding
electron/positron and radiation conductivities $K_e$ and $K_r$ are
\begin{equation}
 K_e = 1.9369 \Bigl(\frac{2}{\pi}\Bigr)^{\frac32} \frac{k^{\frac72}
 T^{\frac52}}{m^{\frac12}_e e^4} \approx 0.98385 \cdot \frac{k^{\frac72}
 T^{\frac52}}{m^{\frac12}_e e^4} \; \; \; , \; \; \;
 K_r = \frac{16 \sigma \lambda_R T^3_r}{3}
\end{equation}
where $\lambda_R$ is the Rosseland mean free path mentioned above. The
electron/positron-radiation coupling constant $A_{er}$ is represented in the
form $A_{er} = C_{ve} (\nu_b + \nu_C)$, where $C_{ve}$ is the specific heat
of the electron-positron plasma, while $\nu_b$ and $\nu_C$ are the overall
bremsstrahlung rate and Compton scattering rate, respectively. In our
radiation diffusion model the Compton scattering rate $\nu_C$ can be
evaluated from the following formula
\begin{equation}
 \nu_C = 134.04 \cdot \frac{\sigma R e^2}{(m_e c^2)^2} \xi T^4_r
\end{equation}
where $R = k N_A$ is the universal gas constant and $\xi$ is the numerical
factor ($\xi \approx 2$). To obtain the exact value of $\xi$ one needs to
take into account an obvious similarity between hydrogen atom H and two-body
positronium system Ps (= $e^- e^+$). Such a similarity can be found for the
bound state spectra of these systems, their continuous spectra,
photodetachment cross-sections, etc. In fact, it can be traced even for
corresponding negatively charged ions H$^-$ and Ps$^-$ each of which has
only one stable bound state ($1^1S$-state \cite{Fro07a}). This means that
for electron-positron plasma we can use the known formula  for the Compton
scattering rate in the pure hydrogen (see, e.g., \cite{Fra}). The only
change in this formula is related to the fact that the Ps ($e^{+} e^{-}$)
system has spatial radius which is twice larger than the analogous spatial
radius of the hydrogen atom H. The same approach can be used to evaluate a
large number of other properties, e.g., the overall bremsstrahlung rate and
Von-Neuman artificial viscosity.

The overall bremsstrahlung rate in macroscopic polyelectrons is written in
the form
\begin{equation}
 \nu_b = \frac{64 \sqrt{2 \pi}}{3} \cdot \frac{\alpha e^2 R^2 \rho
 \xi^{\beta}}{k \sqrt{m_e k T}} \cdot G\Bigl(\frac{T_r}{T_e}\Bigr) \approx
 0.390224 \cdot \frac{e^2 R^2 \rho \xi^{\beta}}{k \sqrt{m_e k T}} \cdot
 G\Bigl(\frac{T_r}{T_e}\Bigr)
\end{equation}
where $\alpha = \frac{e^2}{2 \pi \hbar c}$ is the fine structure constant
and $G(x)$ is the so-called universal bremsstrahlung function
\begin{equation}
 G(y) = \frac{1}{y - 1} \int^{\infty}_0 \frac{dx f(x) \Bigl\{1 -
 exp[\frac{x (y-1)}{y}]\Bigr\}}{1 - exp(-\frac{x}{y})}
\end{equation}
where the function $f(y)$ represents the bremsstrahlung emission spectrum.
Its explicit form is
\begin{eqnarray}
 f(y) &=& \int^{\infty}_1 ln(\sqrt{z} + \sqrt{z+1}) exp(-y z) dz = exp(-y)
 \times \\
 & & \int^{\infty}_0 ln(\sqrt{z+1} + \sqrt{z+2}) exp(-y z) dz =
 \frac{exp(-\frac52 y)}{2 y} I_0\Bigr(\frac{y}{2}\Bigl) - \frac{1 +
 \sqrt{2}}{y} exp(-y) \nonumber
\end{eqnarray}
where $I_0(x)$ is the modified Bessel function (see, e.g., \cite{Wats}).
Note that, if $T_r \gg T_e$, then $\frac{T_r}{T_e} \rightarrow 1, G(x)
\rightarrow 1$ and $\nu_b$ is the pure bremsstrahlung rate. In the
opposite case, i.e. when $T_r \gg T_e$, then $\frac{T_r}{T_e} \rightarrow
\infty, G(x) \rightarrow \frac{\pi^2}{4}$ and $\nu_b$ is the rate of
inverse bremsstrahlung \cite{Fra}.

The equations Eq.(\ref{e2X}), Eq.(\ref{e4X}) and Eq.(\ref{e5X}) govern the
time-evolution of macroscopic polyelectrons. Such a time-evolution also
includes annihilation of the electron-positron pairs in polyelectrons. Our
main hypothesis is related to the fact that at some conditions the
propagation of electron-positron annihilation will be described as a
propagation of moving surfaces in dense polyelectrons. In these cases we can
observe the photon detonation. Currently, we cannot prove that such a
detonation can be detected at realistic densities of the compressed
polyelectrons $\rho \le 1$ $g \cdot cm^{-3}$. At very large densities,
e.g., $\rho \approx 8.5 \cdot 10^2 $ $g \cdot cm^{-3}$, we have, in fact, a
sudden $X-$ray flush of annihilation $\gamma-$quanta, i.e. the annihilation
explosion at high densities. In this case any thermal equilibrium between
particles and radiation cannot be reached. With our equilibrium code
described above, it is hard to understand all details of the transition to
the non-equilibrium process. At this moment our non-equilibrium code written
for electron-positron plasmas does not work due to some unsolved
computational problems. In addition to this, such high densities in
polyelectrons ($\rho \ge 1 \cdot 10^{3} $ $g \cdot cm^{-3}$) are comparable
with the corresponding Fermi limit (see below). Polyleptons compressed to
such densities must be considered as degenerated Fermi systems. This
drastically complicates annihilation analysis in highly compressed
polyelectrons. Nevertheless, some predictions can be made in such cases too,
and we hope to report the results of our analysis in future works.

\section{The low-temperature Fermi limit. Statistics of particles in
polyelectrons.}

Let us briefly describe the Fermi degeneration of identical fermions
(electrons and/or positrons) at high densities and relatively low
temperatures. Such a degeneration is crucial for evaluation of the
feasibility of various explosive devices based on extremely high
compressions (see, e.g., \cite{Fra}). As is well known (see e.g.
\cite{LLS}), if the density of the cold matter is quite high, then all
electrons should be considered as degenerated fermions. In this case the
internal electron pressure $P_e$ as well as the speed of sound $S_s \approx
\sqrt{P_e}$ in a cold electron plasma can suddenly increase to extremely
large values. In highly-compressed polyelectrons we always have both
electrons and positrons; and the overall pressure related to the
electron/positron degeneracy reaches very large values. The appropriate
disassembly time, $\tau \approx \frac{R}{S_s}$, becomes very short and
annihilation in these polyelectrons cannot proceed effectively. Here $R$
means the spatial radius of the highly compressed volume in the
electron-positron plasma, while $S_s$ is the speed of sound. In fact, the
correction on the electron/positron degeneracy is needed when the real
temperature $T$ is comparable with the so-called equivalent Fermi
temperature $T_{ef}$ \cite{LLS}. For the equimolar electron-positron plasmas
the explicit expression for the equivalent Fermi temperature $T_{ef}$ (in
$keV$) takes the form
\begin{eqnarray}
 T_{ef}(keV) = \frac{2}{5} \Bigl(\frac{3 N_A M_p}{8 \pi
 m_e}\Bigr)^{\frac{2}{3}} \cdot \frac{2 \pi^2 \hbar^2}{m_e}
 \rho_0^{\frac{2}{3}} \approx 1.39481 \cdot \rho_0^{\frac{2}{3}}
 \label{fermi}
\end{eqnarray}
where $N_A$ is the Avogadro number, $\hbar$ is the Planck constant, $M_p$ is
the proton mass and $m_e$ electron mass, respectively \cite{COD}. Also, in
this equation $\rho_0$ is the macroscopic density of the electron-positron
plasma in $g \cdot cm^{-3}$. As follows from Eq.(\ref{fermi}) for $T = 0.1$
$keV$ the pressure of the degenerated electron-positron mixture will be
$\approx 14$ times larger than its pressure evaluated with the use of
formulas for an ideal gas.

Note also that the actual statistics of particles in macroscopic
polyelectrons and/or in highly compressed electron-positron mixtures is
still an open question. Based on `atomic analogies' one can expect that both
electrons and positrons must be considered as the two different types of
particles. But, this picture is realistic only for very small,
non-relativistic energies. Moreover, applications of these `atomic
analogies' to real polyelectrons often produce various contradictions with
reality even for non-relativistic polyelectrons. The common reason of such
contradiction is clear; since all electrons and positrons in polyelectrons
must be considered as the two states of the same particle \cite{FrWa}. This
generates a special `exchange' interaction between particles in
polyelectrons. Due to such an exchange interaction all polyelectrons with
equal (and even) numbers of positrons and electrons have remarkable
stability. For instance, the ground ${}^1S-$state in the bi-positronium
Ps$_2$ is very stable. Moreover, all properties of this state are invariant
under charge conjugation ${\cal C}$, i.e., when $e \leftrightarrow -e$. The
bi-positronium ion Ps$_2 e^-$ is not invariant under charge conjugation,
since ${\cal C}$ (Ps$_2 e^-$) = Ps$_2 e^+ \ne $Ps$_2 e^{-}$.

Now, consider the actual statistics of particles in the bi-positronium ion
Ps$_2 e^-$. By applying `atomic analogies' mentioned above one easily finds
that the ground state in this system is the doublet state with the two
independent spin functions
\begin{eqnarray}
 \chi_1 &=& (\alpha_e \beta_e \alpha_e - \beta_e \alpha_e \alpha_e)
 (\alpha_p \beta_p - \beta_p \alpha_p) \\
 \chi_2 &=& (2 \alpha_e \alpha_e \beta_e - \beta_e \alpha_e \alpha_e -
 \alpha_e \beta_e \alpha_e) (\alpha_p \beta_p - \beta_p \alpha_p)
\end{eqnarray}
where the indexes $e$ and $p$ designate electrons and positrons,
respectively, while the notations $\alpha$ and $\beta$ mean the one-electron
spin-up and spin-down functions, respectively. The corresponding (spatial)
projectors for the spatial parts of these two wave functions can be
constructed explicitly, e.g., by using our method developed in \cite{JETP}.
The result of bound state computations with such trial wave functions is
negative, i.e. the ground state in the Ps$_2 e^-$ ion is not bound. In fact,
in this approach the Ps$^-$ (and Ps$^+$) ion and bi-positronium Ps$_2$ are
the only stable polyelectrons in which the total number of particles exceeds
two.

Another approach to determine the statistics of particles in the
bi-positronium ion Ps$_2 e^-$ follows from numerical calculations of the
Ps$_2$ system. It was noticed in numerous calculation that one can rapidly
decrease the variational energy of the Ps$_2$ system by introducing some
special `exchange' symmetry between positrons and electrons. In some works
such a symmetry was called by the charge-conjugation symmetry. If four
particles in the Ps$_2$ system are designated by numbers 1 ($e^-$), 2
($e^-$), 3 ($e^+$) and 4 ($e^+$), then the total wave function of the ground
(singlet) $1^1S-$state can be written in the form
\begin{equation}
 \Psi({\rm Ps}_2) = \frac18 (1 + {\cal{P}}_{13}) (1 + {\cal{P}}_{12})
 (1 + {\cal{P}}_{34}) \phi(1, 2, 3, 4) = \frac12 (1 + {\cal{P}}_{13})
 \Phi(1, 2, 3, 4) \label{Ps2sym}
\end{equation}
where $\phi(1, 2, 3, 4)$ is the non-symmetrized wave function of the
four-particles, while $\Phi(1, 2, 3, 4)$ is the wave function of the Ps$_2$
system symmetrized in respect with `atomic analogies'. The wave function
$\Psi({\rm Ps}_2)$ has an additional symmetry in comparison with the
$\Phi(1, 2, 3, 4)$ function and it allows one to produce very low total
energies for the Ps$_2$ system.

Based on these results for the Ps$_2$ system we can split the bi-positronium
ion Ps$_2 e^-$ into two systems: the central cluster Ps$_2$ and one electron
$e^-$ which moves in the field of this cluster. The cluster Ps$_2$ is a
charge-conjugate system with the wave function symmetrized according in
respect with Eq.(\ref{Ps2sym}). Now, all particles in the Ps$_2$ system
become absolutely identical, while the outer-most electron is a
particle with different permutation symmetry. The result of bound state
computations with trial wave functions symmetrized in respect with this
picture indicate clearly that the ground state in the Ps$_2 e^-$ ion is
bound (in fact, it is well bound \cite{FrWa}). Moreover, it can be shown
that in this approach the ground state in the six-particle polylepton Ps$_2
e^- e^+$ (or Ps$_3$) is also bound. In fact, if the total wave functions are
symmetrized in respect to this approach, than many other polyleptons are
also bound. Furthermore, the total and binding energies of the bound
(ground) states in all bound polyleptons show smooth dependence upon the
number of particles (or number of `electrons'). Below, such an approach to
symmetrization of polyelectrons is called the united statistics.

The Ps$_2 e^-$ ion is only one example of systems for which the united
statistics produces results which contradict statistics based on `atomic
analogies'. Formally, this united statistics must be applied for all
polyelectrons. For some polyelectrons, e.g., for Ps$^-$ and Ps$_2$, such new
statistics leads to the same permutation symmetries of the total wave
functions which is already known from `atomic analogies'. However, in higher
polyelectrons, e.g., in Ps$_2 e^-$, Ps$_3$, etc, the united statistics
produces the wave functions of the unusual symmetry (from the atomic point
of view). In particular, if electrons and positrons in polyelectrons are
allowed to form some very stable bosonic clusters (which are invariant in
respect to charge conjugation), e.g., bi-positronium Ps$_2$, then this will
change the overall interparticle statistics (for more details, see
discussion in \cite{FrWa}).

\section{Production of the electron-positron pairs at high-temperature
explosions.}

In general, an intense creation of electron-positron pairs by high-energy
photons (or $\gamma-$quanta) around every physical `body' heated to
extremely high temperatures, e.g., $T \ge 5 - 10$ $MeV$, can be considered
as an instability of the real media for propagating photons of high and very
high energies. The total numbers of the newly created electrons $N_{-}$ and
positrons $N_{+}$ rapidly increase with the temperature $T$ and can be
evaluated with the use of the following formula (see, e.g., \cite{LLS})
\begin{eqnarray}
 N_{+} \approx N_{-} \approx \frac{3 \zeta(3)}{2 \pi^2} \cdot V \cdot
 \Bigl( \frac{k T}{\hbar c} \Bigr)^3 \label{electr}
\end{eqnarray}
where $V$ is the volume in which positrons are created and $\zeta(x)$ is the
Riemann $\zeta-$function \cite{AS} ($\zeta(3) \approx$ 1.202056903159594).
In fact, this number is small (and even very small) for $k T \ll m_e c^2$,
but for $k T \approx m_e c^2$ the total number of positrons $N_{+}$ in
$\approx 10^6 - 10^7$ times exceeds the numbers of incident atomic electrons
located in the same volume $V$. At temperatures $k T \approx m_e c^2$ each
newly created positron occupies the volume $V_0 \approx \lambda_e = \alpha
a_0$, where $\lambda_e = \frac{\hbar}{m_e c} \approx 3.862 \cdot 10^{-13}$
$m$ is the electron Compton wavelength, $\alpha$ is the fine structure
constant and $a_0$ is the Bohr radius. This means that inside of the volume
of U atom one finds almost $8 \cdot (\alpha)^{-3} \approx 2.057 \cdot 10^7$
positrons (and electrons) created due to the instability of the
electromagnetic vacuum. Note that the total number of atomic electrons in
the neutral U atom equals 92. This means that at such high temperatures,
i.e. for $T \ge 511$ $keV$, we can neglect by all original atomic electrons
and evaluate the total number of electrons by the same formula,
Eq.(\ref{electr}).

Let us suppose that somehow we could create an overheated spatial area with
the radius $\approx 1$ $m$. If the temperature of that area equals $T =
\frac{m_e c^2}{k} \approx 5.93 \cdot 10^9$ $K$, then this overheated ball
contains $\approx 7.272 \cdot 10^{37}$ electrons and positrons. The sudden
annihilation of these $(e^{-},e^{+})-$pairs produces $\approx 1.434 \cdot
10^{38}$ $\gamma-$quanta. The total energy released during this annihilation
explosion equals to thermal explosion of $\approx 1.564 \cdot 10^{12}$
tonnes of TNT, or $1.564 \cdot 10^{6}$ megatonnes of TNT. Such an energy
significantly exceeds thermal energy released during any nuclear and/or
thermonuclear explosion. If the radius of instability area is 1 $cm$, then
annihilation of all electron-positron pairs created in that area produces
thermal energy which is equal to the energy released from an explosion of
1.564 megatonnes of TNT. It can be a very interesting direction to design
various nuclear and thermonuclear explosive devices in which some
macroscopic spatial areas are heated to very large temperatures, e.g., $T
\approx$ 250 - 300 $keV$ and higher.

Creation and following annihilation of large numbers of electron-positron
pairs (or positrons, for short) during nuclear and thermonuclear explosions
has been noticed since the middle of 1950's. The approximate evaluations
made in that time indicated that $\approx$ 5 \% of the total energy of a
typical nuclear explosion is released in the form of newly created
electron-positron pairs. In modern multi-shell nuclear explosive devices
with very light thermonuclear amplifiers the overall and local temperatures
can be much higher. Therefore, the total numbers of positrons (and
electrons) created during such explosions are also much higher. In general,
the local numbers of created positrons/electrons ($dN_{+}$ and/or $dN_{-}$)
directly depend upon the local temperature $T$ (in $eV$) and volume $dV$ of
the spatial area. The exact formula takes the form
\begin{eqnarray}
 dN_{+} = dN_{-} = \frac{dV}{\pi^2 \hbar^3} \int_{0}^{\infty} \frac{p^2
 dp}{exp\Bigl(\frac{E}{T}\Bigr) + 1} \label{number}
\end{eqnarray}
where $E = c \sqrt{p^2 + m^2 c^2}$ is the relativistic energy of a particle
with momentum $p$. This expression must be also integrated over the spatial
volume of the high-temperature area inside of the nuclear explosive device.

Another direction which can be used to reach extremely high temperatures is
based on the use of thermonuclear explosions. For instance, the temperature
developed during a thermonuclear explosion of modern 550 kt (of TNT)
warhead with the total mass 280 kg can be evaluated as
\begin{equation}
 T \approx \frac{550,000,000,000,000 \cdot 8.617343 \cdot 10^{-5}}{280,000}
 \approx 169.3 \cdot 10^3 eV = 169.3 \; \; \; keV \label{temp}
\end{equation}
where we have assumed that 1 gram of TNT releases the energy 1000 $cal$
during its explosion (see, e.g., \cite{Yamp}). The temperature $T$ which is
determined by Eq.(\ref{temp}) corresponds to the case when complete thermal
equilibrium is established between remains of the warhead and radiation. The
local temperatures achieved inside of this explosive device can be
significantly higher ($T \approx$ 300 $keV$). Such temperatures are
comparable with the corresponding threshold energy $\approx$ 511 $keV$ which
is needed for creation of one positron $e^{+}$. In general, for $T < 511$
$keV$ the intensity of annihilation $I_A$ in the electron-positron plasma
per unit volume is
\begin{equation}
 I_A = 4 \pi \alpha^4 \Bigl(\frac{c}{a_0}\Bigr) \Bigl( \frac{m
 T}{2 \pi \hbar^2} \Bigr)^3 \exp\Bigl(- \frac{2 m_e c^2}{T} \Bigr)
 \label{prod}
\end{equation}
where $T$ is the temperature expressed in $eV$. For $T \approx 511$ $keV$
the factor $\Bigl( \frac{m T}{2 \pi \hbar^2} \Bigr)^3 \approx \frac{1}{8
\pi^3} \Bigl( \frac{m c}{\hbar} \Bigr)^6 \sim \alpha^{-6}$, i.e. it is a
very large number. The factor $I_A$ defined in Eq.(\ref{prod}) is the
intensity of electron-positron annihilations observed in one atomic volume
$V_a = \frac{4 \pi}{3} (a_0)^3$.

For regular nuclear and thermonuclear explosions we always have $T \ll 511$
$keV$. In such cases the densities of the electrons $n_{-}$ and positrons
$n_{+}$ at equilibrium can be evaluated from the formulas (see, e.g.,
\cite{LLS})
\begin{eqnarray}
  n_{-} = \frac{n_0}{2} + \Bigl[ \Bigl(\frac{n_0}{2}\Bigr)^2 +
  \frac12 \Bigl(\frac{m T}{2 \pi \hbar^2} \Bigr)^3 \exp\Bigl(- \frac{2 m_e
  c^2}{T} \Bigr) \Bigr]^{\frac12} \label{equ20} \\
  n_{+} = - \frac{n_0}{2} + \Bigl[ \Bigl(\frac{n_0}{2}\Bigr)^2 +
  \frac12 \Bigl(\frac{m T}{2 \pi \hbar^2} \Bigr)^3 \exp\Bigl(- \frac{2 m_e
  c^2}{T} \Bigr) \Bigr]^{\frac12} \label{equ21}
\end{eqnarray}
where $n_0$ is the incident electron density. The formulas, Eqs.(\ref{prod})
- (\ref{equ21}), can be applied to evaluate the actual positron/electron
density and determine the intensity of annihilation $I_A$ in the
electron-positron plasma which forms during any nuclear and/or thermonuclear
explosion. Our analysis indicates clearly that formation and annihilation of
very large numbers of electron-positron pairs inside of modern nuclear and
thermonuclear explosive devices is an important part of the explosion.
Briefly, we can conclude that macroscopic polyelectrons extensively form
during high-temperature thermal explosions, e.g., during any nuclear and/or
thermonuclear explosion. The following annihilation of large numbers of
electron-positron pairs can change the observed spectrum of soft
$\gamma-$radiation emitted in such explosions.

In applications to astrophysics it is important to note that the area in
which electromagnetic vacuum is unstable arises around each overheated
physical body, e.g., around any star with very high temperature at its
surface. Very large numbers of electrons and positrons are constantly
created in such areas. In addition to thermal instability of electromagnetic
vacuum there are many other ways in which positrons (and electrons) are also
created. Later all these newly created particles begin to propagate into
outer space and annihilate during their propagation. In general,
annihilation of electron-positron pairs proceeds very intensively in spatial
areas closed to the overheated body. At larger distances the overall
intensity of the annihilation rapidly decreases with the distance. In
general, these two spatially separated areas (area of vacuum instability and
annihilation area) can be found around any overheated physical body.
Moreover, the spatial radii of the both areas (instability and annihilation
areas) rapidly increase with the temperature of the overheated body.

\section{Thermal support of thermonuclear burn-up.}

Let us describe the problem in which thermal energy released from
electron-positron annihilation can be used to simplify thermonuclear
ignition. For simplicity, consider thermonuclear ignition of the compressed
$\rho \ge 10$ $g \cdot cm^{-3}$ equimolar deuterium-tritium mixture (below,
DT-mixture, for short). As follows from numerous experiments and various
theoretical evaluations after ignition in one spatial point (also called the
central point) such a mixture burns very effectively by itself (see, e.g.
\cite{Fro00}). Practically in all actual cases the burning wave propagates
from the hot center as a high-temperature detonation wave. In reality, the
compressed equimolar DT-mixture can be ignited by an incident nuclear charge
either directly, i.e. by a propagating heat wave, or by using the phenomenon
of radiative ablation. In the last case the ablating radiation flux comes
from from an incident nuclear explosion. The principal question is: can we
ignite a highly compressed equimolar DT mixture without any preliminary
nuclear explosion? In general, to achieve thermonuclear ignition one needs
to compress the equimolar DT mixture to very large densities $\rho \ge$ 500
$g \cdot cm^{-3}$. If such a highly compressed equimolar DT mixture is
confined, e.g., by a convergent consequence of shook waves, for some time
(e.g., $\tau \ge 1 \cdot 10^{-9}$ $sec$), then it burns up by itself. In
actual systems such a confined time must be even longer, since it is very
hard to exclude all possible energy losses.

The current answer to the question formulated above is negative, since at
this time it is impossible to reach such high densities and relatively long
confinement times for the equimolar DT mixture by using only chemical
explosives. This means that the shock wave which propagates from the hot
center of highly compressed equimolar DT-mixture has temperatures which are
not sufficient for thermonuclear ignition of this mixture. This situation
can be changed, in principle, by saturating the compressed DT-mixture with
positrons. The following annihilation of electron-positron pairs and
absorption of annihilation $\gamma-$quanta in the dense DT-mixture can be
used as an additional thermal source which can amplify and accelerate the
shock wave propagating from hot center. The amplified and accelerated shock
wave can produce thermonuclear ignition.

The governing equation takes the form (see, e.g., \cite{AvrFeo})
\begin{equation}
 {\cal C} \cdot \frac{d T}{dt} = - {\cal C} \cdot \frac{3}{r_{f}}
 \cdot \frac{d r_{f}}{dt} \cdot T + \tilde{q}(r_{f},T,\rho_0) =
 - {\cal C} \cdot \frac{3}{r_{f}} \cdot V_{\max} \cdot T +
 \tilde{q}(r_{f},T,\rho_0) \label{burnup}
\end{equation}
where $T$ is the temperature in the hot zone and ${\cal C}$ is the specific
heat (in $MJ \cdot g^{-1} \cdot keV^{-1}$, where $1 MJ = 1 \cdot 10^{6} J$)
per unit mass of thermonuclear fuel. In the first approximation the specific
heat ${\cal C}$ does not depend upon $T$. In fact, such an approximation has
a very good accuracy if: (1) $T \geq 0.1$ $keV$, and (2) the thermonuclear
fuels contains only light elements (with $A \leq 20$ and $Z \leq 10$). In
the last equation $\tilde{q}(r_{f},T,\rho_0)$ is the so-called energy
release function (per unit volume).

In general, Eq.(\ref{burnup}), represents the burn-up problem in its
classical form \cite{AvrFeo}. Let $r_{f}(t) \geq r_0$ be the radius of the
hot, spherical spot created by the shock (thermal) wave to the time $t$. The
velocity of the hot zone expansion is $\frac{d r_{f}}{dt}$, respectively. In
reality, the hot zone expands either by the high-temperature thermal wave
$V_t = (\frac{d r_{f}}{dt})_t$, or by the detonation wave $V_d = (\frac{d
r_f}{dt})_d$. The actual (or observed) velocity of the hot zone expansion
$V_{\max}$ is largest of the two corresponding velocities, i.e. $V_{\max} =
max \Bigl[(\frac{d r_f}{dt})_d, (\frac{d r_f}{dt})_t \Bigr]$. The
temperature behind the shock (or thermal) wave $T = T_{f}$ is significantly
larger than the initial temperature before such a wave, where $T_0 \approx
0$. Now, by introducing the burn-up parameter $x = \rho_{0} \cdot r_{f}$,
one can re-write the burn-up equation, Eq.(\ref{burnup}), to the the form
(see \cite{AvrFeo}):
\begin{equation}
\frac{d T}{dx} = - \frac{3}{x} \cdot T + \frac{q(x,T,\rho_0)}{{\cal C} \cdot
V_{\max}} \label{burn2}
\end{equation}
where $q(x,T,\rho_0) = \frac{\tilde{q}(r_{f},T,\rho_0)}{\rho_{0}}$ is the
so-called normalized energy release function (per unit mass). In contrast
with $r_f$, the burn-up parameter $x$ does not depend significantly on
$\rho_0$. The explicit expressions for the $q(x,T,\rho_0)$ function and for
the ${\cal C}$ and $V_{\max}$ values depend significantly on the ionic
contents of the considered thermonuclear fuel.  For equimolar DT-mixture the
explicit form of the $q(x,T,\rho_0)$ function is very well known (see, e.g.,
\cite{Fra}, \cite{AvrFeo}).

Numerical solution of the burn-up equation, Eq.(\ref{burn2}), is
straightforward, if the function $q(x,T,\rho_0)$ is known. For given density
$\rho_0$ and temperature $T$ this equation allows one to determine the
minimal value of the burn-up parameter $x_{cr}$. Solution of
Eq.(\ref{burn2}) for different temperatures produces the explicit dependence
of minimal (or critical) burn-up parameter upon the temperature $T$. In
other words, for any given density $\rho$ of the equimolar DT-mixture we
obtain the burn-up curve $x_{cr}(T)$. As follows from the results of
earlier works (see, e.g., \cite{Fro00}, \cite{AvrFeo}) the minimal burn-up
parameter rapidly decreases with the temperature. Briefly, this means that
some under-critical equimolar DT-mixture can be transformed into
over-critical DT-plasma (with the same density) by increasing the central
temperature. Here the over-critical DT-plasma means the high-temperature
equimolar DT-plasma in which thermonuclear burn-up is possible. The idea to
use an additional thermal support for thermonuclear burn-up is based on this
fact. If the equimolar DT-mixture is saturated by positrons, then the
thermal energy released from electron-positron annihilation can be used to
simplify thermonuclear ignition. This approach can be used in some
applications, including modern experiments which are based on laser-driven
ablative implosion scheme to compressed small DT-microspheres.

\section{Conclusions.}

We have considered the phenomena of annihilation explosions and photon
detonation in macroscopic polyelectrons. In general, the macroscopic
polyelectrons can be created experimentally with the use of radiative
ablation. Theoretical and experimental study of such polyelectrons is of
great interest in a number of applications. In general, the macroscopic
properties of polyelectrons are different from analogous properties of
regular (i.e. atomic) systems. A very fast compression of macroscopic
polyelectrons creates conditions at which almost sudden annihilation of very
large number of electron-positron pairs is possible. This phenomenon is
called the photon explosion. It can be used to develop various explosive
devices based on electron-positron annihilation. Moreover, in the dense
electron-positron plasma, annihilation of particles can propagate as a
spatial surface which, in fact, coincides with the detonation wave. In such
cases we deal with the photon detonation. Currently, there are a number of
experimental restrictions which are crucial for workability of devices based
on photon explosions. In the future we might expect that the experimental
situation with macroscopic polyelectrons will be improved.

\end{document}